\documentclass[%
 reprint,
superscriptaddress,
 amsmath,amssymb,
 aps,
longbibliography
]{revtex4-1}

\usepackage{graphicx}
\usepackage{dcolumn}
\usepackage{bm}
\usepackage{physics}
\usepackage{mathrsfs}

\begin{document}

\title{Tight finite-key analysis for generalized high-dimensional quantum key distribution}

\title{Tight finite-key analysis for generalized high-dimensional quantum key distribution}
\author{Rong Wang}
\author{Zhen-Qiang Yin}
\email{yinzq@ustc.edu.cn}
\author{Hang Liu}
\author{Shuang Wang}
\author{Wei Chen}
\author{Guang-Can Guo}
\author{Zheng-Fu Han}
\affiliation{CAS Key Laboratory of Quantum Information, CAS Center for Excellence in Quantum Information and Quantum Physics, University of Science and Technology of China, Hefei 230026, China}
\affiliation{State Key Laboratory of Cryptology, P. O. Box 5159, Beijing 100878, P. R. China}





\begin{abstract}
Due to the capability of tolerating high error rate and generating more key bits per trial, high-dimensional quantum key distribution attracts wide interest. Despite great progresses in high-dimensional quantum key distribution, there are still some gaps between theory and experiment. One of these is that the security of the secret key heavily depends on the number of the emitted signals. So far, the existing security proofs are only suitable in the case with an infinite or unpractically large number of emitted signals. Here, by introducing the idea of "key classification" and developing relevant techniques based on the uncertainty relation for smooth entropies, we propose a tight finite-key analysis suitable for generalized high-dimensional quantum key distribution protocols. Benefitting from our theory, high-dimensional quantum key distribution protocols with finite resources become experimentally feasible.
\end{abstract}

\pacs{Valid PACS appear here}
\maketitle


\section{\label{sec:level1}INTRODUCTION}

\noindent Quantum key distribution (QKD), considered as the first application in quantum information science, can provide two distant parties Alice and Bob with a string of secret key bits by the laws of quantum mechanics. Because of this amazing feature, it has been rapidly developed in both theory and experiment over the last three decades \cite{BB84,ekert1991quantum,shor2000simple,gobby2004quantum,renner2008security,lo2012measurement,braunstein2012side,sasaki2014practical,pirandola2017fundamental,lucamarini2018overcoming,boaron2018secure,wang2019beating}. Among all the proposed QKD protocols, most of them are based on qubit systems, such as the well-known BB84 protocol \cite{BB84}. QKD protocols using qubit systems are very mature both in theory and experiment, but in some scenarios their performances are limited due to the dimensionality. For instance, each qubit can distribute at most one key bit. As our requirements for protocol performance increase, more and more novel protocols have been proposed. Some of them can tolerate high error rate such as six-state protocol \cite{bruss1998optimal}, some of them carry more than one secret key in each signal \cite{cerf2002security}. Some of these QKD protocols prepare quantum states in a Hilbert space larger than $2$, while others may prepare and measure quantum states in two or more bases. That is the reason we call them high-dimensional(HD) QKD protocols.

Since HD-QKD has various advantages, scholars have made a lot of efforts both in its security proofs and in experimental techniques \cite{cerf2002security,ali2007large,walborn2006quantum,mirhosseini2015high}. However, the existing security proofs \cite{cerf2002security,sheridan2010security,yin2018improved} are only available under the assumption that we have infinite resources. In other words, the two parties Alice and Bob are required to exchange arbitrarily large quantum signals $N$, which cannot be achieved by practical equipment. When we remove the infinite resources assumption, that is, when we consider the finite-key issue, several security proofs \cite{sheridan2010security,sheridan2010finite,bradler2016finite} have been proposed for some specific HD-QKD protocols. Frustratingly, the number of exchanged quantum signals $N$ is usually too large to be realized. Thus, a more efficient method to reduce $N$ to an acceptable level is an urgent need. Additionally, the existing proofs  \cite{tomamichel2012tight} for HD-QKD protocols are not general, e.g., Bob is assumed to make measurements along only two bases albeit coding states are qudit systems.  

Here, we propose an efficient method to tackle finite-key issues for generalized HD-QKD protocols, i.e. the dimension of Hilbert space is arbitrary and Bob's measurement bases can be multiple. The proposed method can cover the previous proof technique \cite{tomamichel2012tight} that is only suitable for two measurements bases. The essential feature of our method is introducing the idea of "key classification", which means classifying key bits into different types with different bit error patterns. Furthermore, applying the uncertainty relation \cite{berta2010uncertainty} for smooth entropies \cite{tomamichel2011uncertainty} to each type and developing relevant theoretical techniques, we derive a tight bound of the secret key rate for HD-QKD in finite-key scenarios. Compared with previous methods including the de Finetti theorem \cite{renner2007symmetry} and postselection technique \cite{christandl2009postselection}, our method leads to a more optimistic bound. Through numerical simulations, we show that, for a variety of HD-QKD protocols, the number of exchanged quantum states $N$ can be reduced dramatically thanks to the proposed theory.

\section{SECURITY DEFINITION}

Before stating our proof technique, let us review the security framework \cite{renner2008security,muller2009composability} that we are concerned about in this paper. A general QKD protocol is executed by two distant parties Alice and Bob. Bob receives the signals from an insecure quantum channel. Then Alice and Bob output either a pair of bit strings $S_{A}$ and $S_{B}$, or a symbol $\bot$ to indicate the abort of the protocol. 

According to the definition of security, a QKD protocol has to satisfy three criteria called "correctness", "secrecy" and  "robustness". Owing to the practical implementation, it is impossible to guarantee $S_{A}=S_{B}$. Then a QKD protocol is $\varepsilon_{cor}$-correct, if it is  $\varepsilon_{cor}$-indistinguishable from a $S_{A}=S_{B}$ protocol. Similarly, a protocol is $\varepsilon_{sec}$-secret, if
\begin{equation}
\frac{1}{2}||\rho_{AE}-U_{A}\otimes\rho_{E}||\le\varepsilon_{sec},
\end{equation}
where $U_{A}$ is the fully mixed state of Alice's system, $\rho_{AE}$ is the composed state of Alice and Eve and $\rho_{E}$ is its reduced density matrix, and $||\cdot ||$ denotes the trace norm. Finally, a protocol is  $\varepsilon_{rob}$-robust, if there exists an honest implementation where the probability that the protocol does not abort is at least $1-\varepsilon_{rob}$. In this work, for simplicity, we just consider the correctness and secrecy of a QKD protocol. Thereby, we say a QKD protocol is $\varepsilon_{tot}$-secure, if it is both $\varepsilon_{cor}$-correct and $\varepsilon_{sec}$-secret, with $\varepsilon_{cor}+\varepsilon_{sec}\le \varepsilon_{tot}$.

\section{NOTATION}

Based on this security definition, we are able to guarantee the security when we use our technique in the HD-QKD protocols.  In this work, we take (d+1)-basis QKD protocols, i.e. the generalization of the six-state protocol, as the examples to introduce our proof technique. In order to clearly describe the protocols, we list some notations and assumptions as follows.

First, Alice controls her devices to prepare  $d$-level ($d$ is a prime number in this work) quantum states (qudits) chosen from $d+1$ mutually unbiased bases (MUBs) $\mathbb{X}_{j, k} \in \{\mathbb{X}_{0, 1}, \mathbb{X}_{1, 0}, \cdots , \mathbb{X}_{1, k}, \cdots , \mathbb{X}_{1, d-1}\}$, where the notions are analogous to \cite{sheridan2010security}.  We recall that there are at most $d+1$ MUBs in the $d$-level Hilbert space. Then, Alice randomly chooses one of the MUBs and encodes the key bit into one of its eigenstates. After Bob receives the particle, he is able to randomly choose one MUB to measure it. 

Second, we review the definition of "overlap". The overlap of any two measurements is defined as $ c=\text{max}_{x,z}||\sqrt{M_{x}}\sqrt{N_{z}}||^{2}_{\infty}$, where $\{M_{x}\}$ and $\{N_{z}\}$ are the elements of the positive operator valued measurements (POVMs) of $\mathbb{M}$ basis and $\mathbb{N}$ basis, respectively. In this paper, we heavily rely on the fact that the overlap of any two POVMs of an MUB in $d$-level Hilbert space is $1/d$.  

Third, there exists an equivalent entanglement-based (EB) protocol according to the model described above. Under the EB version of protocol, Alice prepares two entangled quantum states and sends one of them to Bob in each trial. At measurement, we assume that Bob is able to delay all the measurements in $\mathbb{X}_{0, 1}$-basis until parameter estimation is completed. This assumption does not affect the final key rate if the measurement statistics is the same as the ones of actual devices. 

Finally, in practical optical schemes, (d+1)-basis QKD protocols are often realized by weak coherent light rather than a single-photon source. And this does not meet the assumption that Alice prepares $d$-level quantum states. 
Inspired by Lim {\it et al.}'s work \cite{lim2014concise}, the finite-key analysis under this case can intuitively be solved by using decoy states \cite{hwang2003quantum,lo2005decoy,wang2005beating,ma2005practical}. 

We now define a family of (d+1)-basis QKD protocols, $\Phi[n, m, l, \varepsilon_{cor}, leak_{EC}]$, where $n$ is the block size with respect to the sifted keys in $\mathbb{X}_{0, 1}$-basis, $m$ is the number of dits used to do parameter estimation with regard to each basis, $l$ is the secret key length, $\varepsilon_{cor}$ is the required correctness, and $leak_{EC}$ is the information leakage in error correction. The protocol is asymmetric, specifically, the $n$ sifted keys used for producing final secret keys are measured in $\mathbb{X}_{0, 1}$-basis, the other $(d+1)*m$ dits used for parameter estimation are measured in all $d+1$ bases. Therefore, the number of total sifted keys is defined as $N=n+(d+1)*m$.
The protocol is described as follows.  

\section{PROTOCOL DESCRIPTION}

State Preparation: Alice and Bob repeat the first four steps of the protocol for $i=1, \cdots , M$ until the condition in the sifting step is met. Alice chooses a basis $\mathit{X}_{i} \in \{\mathbb{X}_{0, 1}, \mathbb{X}_{1, 0}, \cdots , \mathbb{X}_{1, k}, \cdots , \mathbb{X}_{1, d-1}\}$, where $\mathbb{X}_{j, k}$ is chosen with probability $p_{j, k}$ respectively. Here we choose $p_{0,1}=f(n, m)$ and $p_{1,k}=(1-p_{0,1})/d$ that the function $f(n, m)$ is chosen to minimize the number $M$ of exchanged quantum states. Then, Alice chooses a random dit $r_{i}\in \{0, 1, \cdots ,d-1\} $ and prepares the quantum state corresponding to $r_{i}$ in a basis $\mathit{X}_{i}$. 

Distribution: Alice sends the quantum state over the insecure channel to Bob.  

Measurement: Bob also chooses a basis $\tilde{\mathit{X}_{i}}\in  \{\mathbb{X}_{0, 1}, \mathbb{X}_{1, 0}, \cdots , \mathbb{X}_{1, k}, \cdots , \mathbb{X}_{1, d-1}\}$ with probability $p_{j,k}$ respectively. After receiving the state, Bob measures it in the chosen basis and stores the outcome $\tilde{r_{i}}\in \{0, \cdots ,d-1\}$. 

Sifting: Alice and Bob broadcast their basis settings over a classical authenticated channel. We define the sets $\mathfrak{X}_{0,1}:= \{i: \mathit{X}_{i}=\tilde{\mathit{X}_{i}}=\mathbb{X}_{0,1}\}$ and $\mathfrak{X}_{1,k}:= \{i: \mathit{X}_{i}=\tilde{\mathit{X}_{i}}=\mathbb{X}_{1,k}\}$. The protocol repeats the first four steps unless $|\mathfrak{X}_{0,1}|\ge n+m$ and $|\mathfrak{X}_{1,k}|\ge m$ for each $k \in \{0, \cdots , d-1\}$. 

Parameter estimation: Alice and Bob use $n$ random dits from $\mathfrak{X}_{0,1}$ to form the code dit strings $\mathbf{X}_{0,1}^{n}$ and $\bar{\mathbf{X}}_{0,1}^{n}$, respectively. Then, for m dits from $\mathfrak{X}_{j,k} \in \{\mathfrak{X}_{0, 1}, \mathfrak{X}_{1, 0}, \cdots , \mathfrak{X}_{1, k}, \cdots , \mathfrak{X}_{1, d-1}\}$, they compute $d$ types of statistical parameters $q^{(t)}_{j,k}:=\frac{1}{m}\sum_{i} \delta^{(t)}_{j,k}$ where 
\begin{equation}
\label{eq6}
\delta^{(t)}_{j,k}=\left\{
\begin{aligned}
1 & , & \tilde{r_{i}}-r_{i} \pmod{d}=t, \\
0 & , & \tilde{r_{i}}-r_{i} \pmod{d} \neq t,
\end{aligned}
\right.
\nonumber
\end{equation}
and $t\in \{{0, 1, 2, \cdots , d-1\}}$. Moreover, these parameters satisfy $\sum_{t=0}^{d-1}q^{(t)}_{j,k}=1$ with the probability of no error $q^{(0)}_{j,k}$ for each basis $\mathbb{X}_{j,k}$. The protocol aborts if the probability of error $\sum_{t=1}^{d-1}q^{(t)}_{j,k}$ for each basis $\mathbb{X}_{j,k}$ is too high. 

Error correction: For those $n$ that pass the parameter estimation step, an information reconciliation scheme is applied. This allows Bob to obtain an estimate $\hat{\mathbf{X}}_{0,1}^{n}$ of $\mathbf{X}_{0,1}^{n}$ by Alice sending him $leak_{EC}$ bits of error correction data. Then, Alice computes a bit string (a hash) of length $\left \lceil \log_{2}\frac{1}{\varepsilon_{cor}} \right \rceil$ by using a random two-universal hash function to $\mathbf{X}_{0,1}^{n}$. She sends the choice of function and the hash to Bob. The protocol aborts if $\text{hash}(\hat{\mathbf{X}}_{0,1}^{n}) \neq \text{hash}(\mathbf{X}_{0,1}^{n})$. 

Privacy amplification: If the $n$ dits pass the error correction, Alice and Bob apply a random two-universal hash function to $\mathbf{X}_{0,1}^{n}$ and $\hat{\mathbf{X}}_{0,1}^{n}$ to extract the final secret $l$ bits ($l*\log_{d}2$ dits)

\section{SECURITY ANALYSIS}

We now present our main result of our paper. It says that the (d+1)-basis protocols $\Phi[n, m, l, \varepsilon_{cor}, leak_{EC}]$ are both $\varepsilon_{cor}$-correct and $\varepsilon_{sec}$-secret, if the length of the secret key is calculated according to a given set of observed values. The correctness is guaranteed by the error correction step, where a hash of Alice's sifted key is compared with the hash of its estimate of Bob. For simplicity, we assume that the quantum channel can be simulated as a generalization of the qubit depolarizing channel which leads to
\begin{equation}
\begin{aligned}
q^{(0)}_{j,k}=1-Q   \quad    q^{(1)}_{j,k}=\cdots = q^{(d-1)}_{j,k}=\frac{Q}{d-1},    
\end{aligned}
\end{equation}
for each basis $\mathbb{X}_{j,k}$. If the length of secret key $l$ satisfies 
\begin{equation}
\begin{aligned}
l \le & n(\log_{2}d-H({\underline{\xi}}))(1-Q-\mu(\varepsilon))\\
       &-leak_{EC}-\log_{2}\frac{2}{\varepsilon^{2}_{sec}\varepsilon_{cor}}, \\     
\end{aligned}
\end{equation}
the protocols $\Phi[n, m, l, \varepsilon_{cor}, leak_{EC}]$ are $\varepsilon_{sec}$-secret. In this formula, $\underline{\xi}$ is a $d$-level probability vector denoted by $\underline{\xi}=\{\xi_{0}, \xi_{1}, \cdots , \xi_{t}, \cdots , \xi_{d-1} \}$, and 
\begin{equation}
\begin{aligned}
\xi_{0}=& \frac{1-\frac{d+1}{d}(Q+\mu(\varepsilon))}{1-Q-\mu(\varepsilon)} \\
\xi_{1}=&\cdots = \xi_{d-1}=\frac{\frac{1}{d(d-1)}(Q+\mu(\varepsilon))}{1-Q-\mu(\varepsilon)}, \\     
\end{aligned}
\end{equation}
where $H(\cdot)$ denotes the entropy function of the $d$-level probability vector by $H({\underline{\xi}})=\sum_{t=0}^{d-1}-\xi_{t}\log_{2}\xi_{t}$, $\varepsilon_{sec}=4\sqrt{1-(1-\varepsilon^{2})^{d+1}}$ and $\mu(\varepsilon)$ that accounts for statistical fluctuation is given by 
\begin{equation}
\mu(\varepsilon):=\sqrt{\frac{n+m}{nm}\frac{m+1}{m}\ln \frac{1}{\varepsilon}},
\end{equation}
(in the following, we will simplify $\mu(\varepsilon)$ as $\mu$). When we comes to the asymptotic case of sufficiently large block sizes $n$, the statistical fluctuation term $\mu$ can be neglected, and thus $l$ satisfies $l \le n(\log_{2}d-H({\underline{\xi}}))(1-Q)-leak_{EC}$, as obtained in previous work \cite{cerf2002security}.

Here we show a sketch of the proof of equation (3), and a rigorous proof including a more general version of the equation (3) can be found in Appendix B. We denote the dit strings of length $n$ by $\mathbf{X}_{0,1}^{n}$ of Alice's side and $\bar{\mathbf{X}}_{0,1}^{n}$ of Bob's side, respectively, which are used to extract the final key. Then, after the measurements (based on EB version), the classical-classical-quantum state of Alice, Bob and Eve is given by
\begin{equation}
\begin{aligned}
\rho_{\mathbf{X}_{0,1}^{n}\bar{\mathbf{X}}_{0,1}^{n}E}=& \sum_{x_{0,1}^{n}, \bar{x}_{0,1}^{n}}\mathbb{P}(x_{0,1}^{n}, \bar{x}_{0,1}^{n})\\
                            & \ket{x_{0,1}^{n}, \bar{x}_{0,1}^{n}}_{AB}\bra{x_{0,1}^{n}, \bar{x}_{0,1}^{n}} \otimes \rho_{E|{x_{0,1}^{n}\bar{x}_{0,1}^{n}}},   \\
\end{aligned}
\end{equation}
where $x_{0,1}^{n}\in\mathbf{X}_{0,1}^{n}$ and $\bar{x}_{0,1}^{n}\in \bar{\mathbf{X}}_{0,1}^{n}$ respectively, and $\mathbb{P}(x_{0,1}^{n}, \bar{x}_{0,1}^{n})$ is the probability of joint dit string ($x_{0,1}^{n}, \bar{x}_{0,1}^{n}$). Then, owing to the error patterns, we define a dit $y_{0,1}^{n} \in \mathbf{Y}_{0,1}^{n}$ that is given by 
\begin{equation}
\begin{aligned}
&y_{0,1}^{n}:=\bar{x}_{0,1}^{n}-x_{0,1}^{n} \pmod{d} \\
\end{aligned}
\end{equation}
where the subtraction is bitwise (it can be considered as a generalization of the XOR operation on dits). Then we "classify" the state $\rho_{\mathbf{X}_{0,1}^{n}\bar{\mathbf{X}}_{0,1}^{n}E}$ according to the dit string $y_{0,1}^{n}$, and define the conditional state
\begin{equation}
\begin{aligned}
\rho_{\mathbf{X}_{0,1}^{n}\bar{\mathbf{X}}_{0,1}^{n}E|y_{0,1}^{n}}=& \sum_{\bar{x}_{0,1}^{n}-x_{0,1}^{n}=y_{0,1}^{n}} \frac{\mathbb{P}(x_{0,1}^{n}, \bar{x}_{0,1}^{n})}{\sum_{\bar{x}_{0,1}^{n}-x_{0,1}^{n}=y_{0,1}^{n}}\mathbb{P}(x_{0,1}^{n}, \bar{x}_{0,1}^{n})}\\
                            & \ket{x_{0,1}^{n}, \bar{x}_{0,1}^{n}}_{AB}\bra{x_{0,1}^{n}, \bar{x}_{0,1}^{n}} \otimes \rho_{E|x_{0,1}^{n}\bar{x}_{0,1}^{n}},   \\
\end{aligned}
\end{equation}
and its corresponding probability by $\mathbb{P}(y_{0,1}^{n})$ which is a marginal probability distribution $\mathbb{P}(x_{0,1}^{n}, \bar{x}_{0,1}^{n})$. Therefore, $\rho_{\mathbf{X}_{0,1}^{n} \bar{\mathbf{X}}_{0,1}^{n}E}$ can be rewritten by
\begin{equation}
\begin{aligned}
\rho_{\mathbf{X}_{0,1}^{n} \bar{\mathbf{X}}_{0,1}^{n}E}= \sum_{y_{0,1}^{n}}} \mathbb{P}(y_{0,1}^{n}) \rho_{\mathbf{X}_{0,1}^{n} \bar{\mathbf{X}}_{0,1}^{n}E|y_{0,1}^{n}. 
\end{aligned}
\end{equation}
For a conditional state $\rho_{\mathbf{X}_{0,1}^{n}\bar{\mathbf{X}}_{0,1}^{n}E|y_{0,1}^{n}}$, we consider two POVMs. One is the $\mathbb{X}_{0, 1}^{\otimes n}$,  the other one is $\mathbb{X}_{1, \vec{j}}^{\otimes n} (y_{0,1}^{n})$ given by
\begin{equation}
\begin{aligned}
\mathbb{X}_{1, \vec{j}}^{\otimes n} (y_{0,1}^{n})=\mathbb{X}_{1, j_{1}} \otimes \cdots \otimes\mathbb{X}_{1, j_{i}} \otimes \cdots \otimes\mathbb{X}_{1, j_{n}}. 
\end{aligned}
\end{equation}
where $j_{i}$ is exactly the $i_{\text{th}}$ dit of dit string $y_{0,1}^{n}$. Then, we use the uncertainty relation for smooth entropies \cite{tomamichel2011uncertainty}. For any tripartite quantum state $\rho_{ABE} \in \mathcal{H}_{ABE}$, the following relation holds
\begin{equation}
\begin{aligned}
&H_{\text{min}}^{\bar{\varepsilon}}(\mathbf{X}_{0,1}^{n}|E)+H_{\text{max}}^{\bar{\varepsilon}}(\mathbf{X}_{1, \vec{j}}^{n} |B) \\
& \ge \log_{2} \frac{1}{c}=n\log_{2}d. \\
\end{aligned}
\end{equation}
In above formula, the smooth min-entropy is for the post-measurement state $\rho_{\mathbf{X}_{0,1}^{n}E|y_{0,1}^{n}}=\text{Tr}_{\text{B}} (\rho_{\mathbf{X}_{0,1}^{n}\bar{\mathbf{X}}_{0,1}^{n}E|y_{0,1}^{n}})$, the smooth max-entropy is for the post-measurement state $\rho_{\mathbf{X}_{1, \vec{j}}^{n}B}$,that is from measuring $\rho_{ABE}$ on POVM $\mathbb{X}_{1, \vec{j}}^{\otimes n} (y_{0,1}^{n})$ and tracing $E$, and we have used the fact that, for each $y_{0,1}^{n}$, the overlap of $\mathbb{X}_{0, 1}^{\otimes n}$ and $\mathbb{X}_{1, \vec{j}}^{\otimes n} (y_{0,1}^{n})$ is $1/d^n$. As we know, there are $d^n$ dit strings in the set $\mathbf{Y}_{0,1}^{n}$, that is, there are $d^n$ smooth min-entropies. With the help of sub-additivity of min-entropy \cite{tomamichel2010duality}, we can connect the smooth min-entropy for $\rho_{\mathbf{X}_{0,1}^{n}E}=\text{Tr}_{\text{B}}(\rho_{\mathbf{X}_{0,1}^{n}\bar{\mathbf{X}}_{0,1}^{n}E})$ and $d^n$ smooth min-entropies for each $\rho_{\mathbf{X}_{0,1}^{n}E|y_{0,1}^{n}}$. Besides, we also bound the sum of $d^n$ smooth max-entropies upon the probability $\mathbb{P}(y_{0,1}^{n})$. Finally, we obtain the lower bound of smooth min-entropy for $\rho_{\mathbf{X}_{0,1}^{n}E}$ given by 
\begin{equation}
\begin{aligned}
H_{\text{min}}^{\tilde{\varepsilon}}(\mathbf{X}_{0,1}^{n}|E) \ge n(\log_{2}d-H({\underline{\xi}}))(1-Q-\mu(\varepsilon)),
\end{aligned}
\end{equation}
where $\tilde{\varepsilon}=\sqrt{1-(1-\varepsilon^{2})^{d+1}}$ and $\mu(\varepsilon)$ that analogously to \cite{tomamichel2012tight} accounts for statistical fluctuation depends on the security parameter $\tilde{\varepsilon}$. The method to bound the sum of smooth max-entropies by observed values, the approach to using sub-additivity of min-entropy and the relation of these parameters $\varepsilon, \bar{\varepsilon}, \tilde{\varepsilon}$ will be found in Appendix B.

Due to the Quantum Leftover Hashing lemma \cite{renner2008security,tomamichel2011leftover}, it is possible to extract a $\Delta$-secret key of length $l$ from $\mathbf{X}_{0,1}^{n}$, where
\begin{equation}
\begin{aligned}
\Delta=2\tilde{\varepsilon}+\frac{1}{2}\sqrt{2^{l-H_{\text{min}}^{\tilde{\varepsilon}}(\mathbf{X}_{0,1}^{n}|E^{'})}}.
\end{aligned}
\end{equation}
The term $E^{'}$ that represents all information Eve obtained can be decomposed as $E^{'}=CE$, where $C$ is classical information revealed by Alice and Bob during the error correction step. For the revealed information that $C$ is at most $leak_{EC}-\log_{2}\frac{2}{\varepsilon_{cor}}$ bits, we use a chain rule for smooth entropies and then obtain
\begin{equation}
\begin{aligned}
H_{\text{min}}^{\tilde{\varepsilon}}(\mathbf{X}_{0,1}^{n}|E^{'}) \ge H_{\text{min}}^{\tilde{\varepsilon}}(\mathbf{X}_{0,1}^{n}|E)-leak_{EC}-\log_{2}\frac{2}{\varepsilon_{cor}}.
\end{aligned}
\end{equation}
If we choose $\tilde{\varepsilon}=\varepsilon_{sec}/4$, combining equation (4) and the Quantum Leftover Hashing lemma, we get 
\begin{equation}
\begin{aligned}
\Delta \le 2\tilde{\varepsilon}+\frac{1}{2}\sqrt{2^{l-H_{\text{min}}^{\tilde{\varepsilon}}(\mathbf{X}_{0,1}^{n}|E^{'})}} \le \frac{\varepsilon_{sec}}{2}+\frac{\varepsilon_{sec}}{2}.
\end{aligned}
\end{equation}
Thus, these protocols are $\varepsilon_{sec}$-secret.


\section{DISCUSSION AND CONCLUSION}

In this section, we analyze the behavior of our security bounds and compare our bounds with previous results by numerical simulations \cite{sheridan2010security}. 
To maximize the expected key rate, we fix $\varepsilon_{tot}=10^{-10}$ and assume an error correction leakage of $leak_{EC}=\zeta nH_{2}(Q+\mu(\varepsilon))$ where $\zeta=1.1$ is the error correction efficiency and $H_{2}(\cdot)$ denotes the binary entropy function.

\begin{figure}[htbp]
\centering
\includegraphics[width=\linewidth]{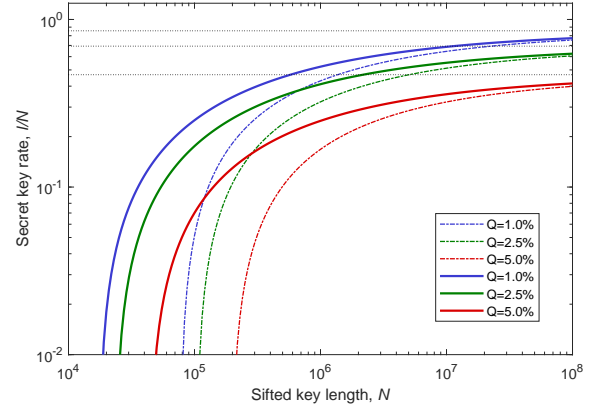}
\caption{The plots show the secret key rate $l/N$ versus sifted key length $N=n+(d+1)*m$ for the protocol when dimension $d=2$ (exactly six-state protocol). The solid curves show our results while the dash-dotted curves show the results given in Ref.\cite{cerf2002security}. The horizontal dashed lines represent the asymptotic rates for error rate $Q\in \{ 1\%, 2.5\%, 5\%\}$ (from top to bottom).}
\label{fig:false-color}
\end{figure}

\begin{figure}[htbp]
\centering
\includegraphics[width=\linewidth]{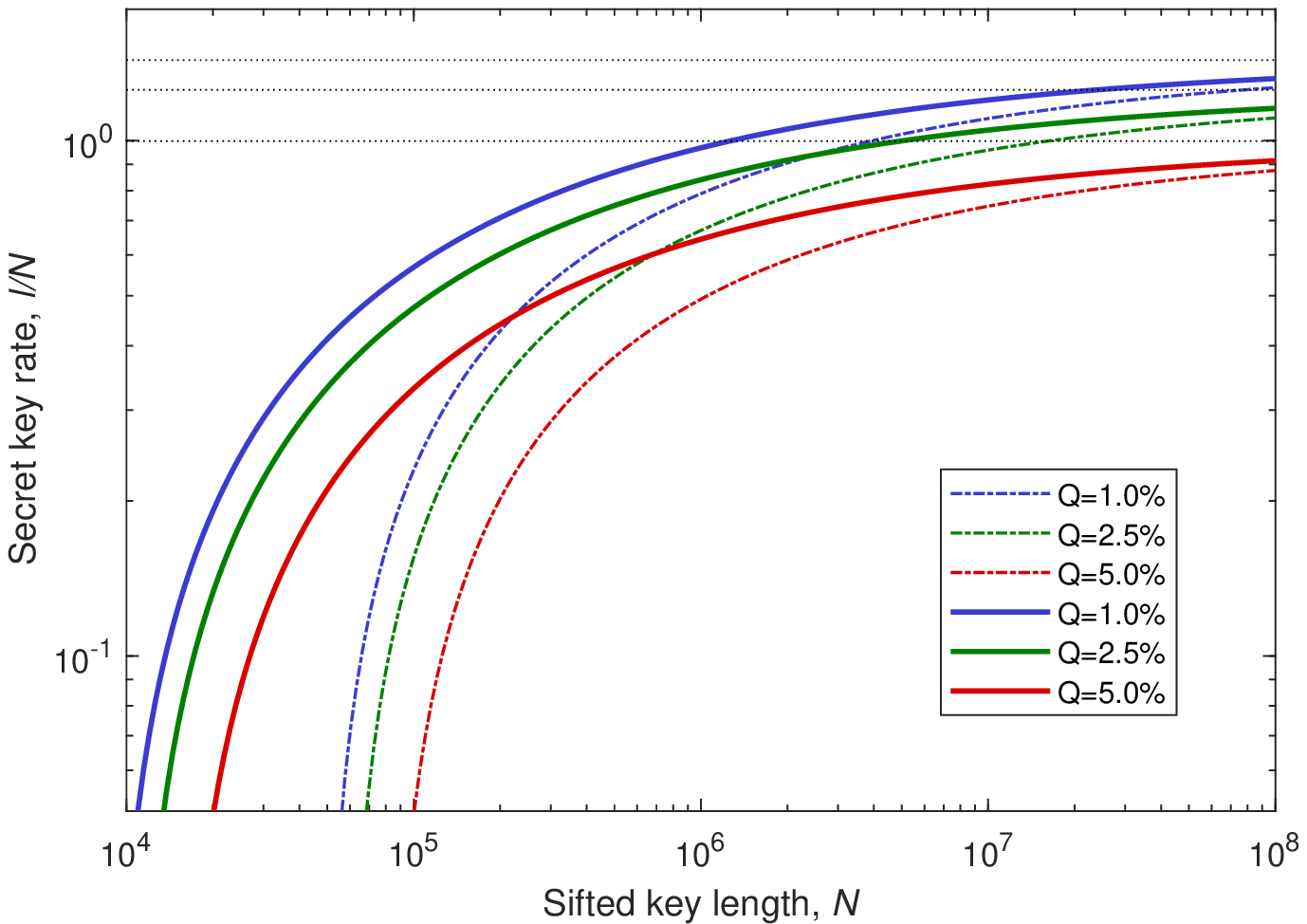}
\caption{The plots show the secret key rate $l/N$ versus sifted key length $N=n+(d+1)*m$ for the protocol when dimension $d=3$. The solid curves show our results while the dash-dotted curves show the results given in Ref.\cite{cerf2002security}. The horizontal dashed lines represent the asymptotic rates for error rate $Q\in \{ 1\%, 2.5\%, 5\%\}$ (from top to bottom).}
\label{fig:false-color}
\end{figure}

\begin{figure}[htbp]
\centering
\includegraphics[width=\linewidth]{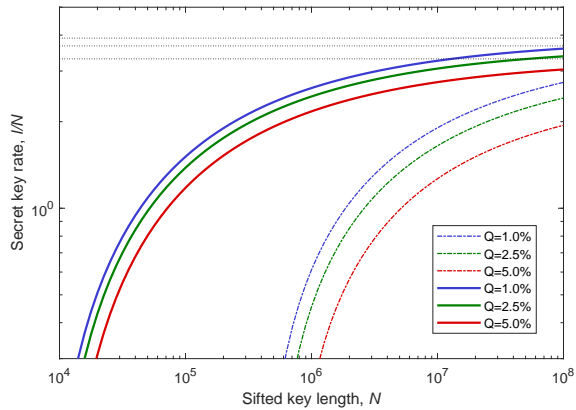}
\caption{The plots show the secret key rate $l/N$ versus sifted key length $N=n+(d+1)*m$ for the protocol when dimension $d=17$. The solid curves show our results while the dash-dotted curves show the results given in Ref.\cite{cerf2002security}. The horizontal dashed lines represent the asymptotic rates for error rate $Q\in \{ 1\%, 2.5\%, 5\%\}$ (from top to bottom).}
\label{fig:false-color}
\end{figure}

In Fig. 1, Fig. 2 and Fig. 3, we compare our optimal key rates (defined as $l/N$) with the secret key rates in \cite{sheridan2010security} of (d+1)-basis QKD protocols featured by $d=2$, $d=3$ and $d=17$ respectively.  As we can see from these figures, our results show a significant improvement in the minimum block size of producing the secret key. Moreover, we can reasonably conjecture that such improvement becomes more prominent with $d$ increasing. Similarly to \cite{tomamichel2012tight}, the improvement is mainly credited to classifying sifted key with error types $y_{0,1}^{n}$ and using entropic uncertainty relation to estimate smooth min-entropy. 

For symmetric (d+1)-basis QKD, that is, when Alice and Bob choose with uniform probability, the key rate may decrease because of discarding the basis mismatch rounds. However, we see from Fig. 1, Fig. 2 and Fig. 3, the asymptotic key rate of asymmetric (d+1)-basis QKD is increasing with dimensionality increasing. It is mainly due to the following reasons. We only consider the number of basis match rounds, that is, the sifted key length. Secondly, as the sifted key length by $N$ is increasing, $p_{0,1}$ (the probability that Alice and Bob choose the key generation basis $\mathbb{X}_{0, 1}$) tends to $1$; therefore, the asymptotic key rate tends to $(\log_{2}d-H({\underline{\xi}}))(1-Q)-H_2(Q)$.

In conclusion, we have given tight finite-key bounds for (d+1)-basis QKD protocols against general attacks. Previous proof techniques cannot effectively tackle multiple measurements QKD protocols such as six-state protocol in the finite-key region. To solve this problem, we propose a proof technique combining a so-called "key classification" idea and entropic uncertainty relation. The "key classification" idea states that we can divide the classical-classical-quantum state $\rho_{\mathbf{X}_{0,1}^{n}\bar{\mathbf{X}}_{0,1}^{n}E}$ into different types according to the relevant dit error patterns, and then apply the entropic uncertainty relation to these states respectively. The subtlety of our proof technique is that we can flexibly classify $\rho_{\mathbf{X}_{0,1}^{n}\bar{\mathbf{X}}_{0,1}^{n}E}$ and construct the corresponding form of the entropic uncertainty relation, which is also the reason that our proof technique can cover the old one \cite{tomamichel2012tight}. Finally, we believe that our proof technique can give more tight finite-key bounds for other high-dimensional QKD protocols such as tomographic \cite{watanabe2008tomography,bouchard2018experimental} and reference-frame-independent \cite{laing2010reference} QKD protocols.


\begin{acknowledgments}
This work has been supported by the National Key Research and Development Program of China (Grant No. 2018YFA0306400), the National Natural Science Foundation of China (Grant Nos. 61822115, 61961136004, 61775207, 61627820) and Anhui Initiative in Quantum Information Technologies.
\end{acknowledgments}


\onecolumngrid
\appendix*
\section*{appendix a: a simple example}
In this section, we take the simplest example with respect to d=2 (exactly the six-state protocol) to introduce our method, and differently from the main text we use some simplified notations. With the assumption of a qubit depolarizing channel, the quantum error bit rates $Q_x$, $Q_y$ and $Q_z$ with respect to $\mathbb{X}$-basis, $\mathbb{Y}$-basis and $\mathbb{Z}$-basis respectively satisfy $Q_x=Q_y=Q_z=Q$. Additionally, following the notations in the main text, $\mathbb{X}$, $\mathbb{Y}$, and $\mathbb{Z}$ are, respectively, $\mathbb{X}_{0,1}$, $\mathbb{X}_{1,0}$, and $\mathbb{X}_{1,1}$. We say a six-state protocol $\Phi[n, m, l, \varepsilon_{cor}, leak_{EC}]$ is $\varepsilon_{sec}$-secret, if the length of secret key $l$ satisfies 
\begin{equation}
\begin{aligned}
l \le  n[1-H_{2}(\frac{1-\frac{3}{2}(Q+\mu)}{1-Q-\mu})][1-Q-\mu]
       -leak_{EC}-\log_{2}\frac{2}{\varepsilon^{2}_{sec}\varepsilon_{cor}},   
\end{aligned}
\end{equation}
where $\varepsilon_{sec}=4\sqrt{1-(1-\varepsilon^{2})^{3}}$. We denote the bit string measured in $\mathbb{X}$-basis of length $n$ by $\mathbf{X}^{n}$ of Alice's side and $\bar{\mathbf{X}}^{n}$ of Bob's side respectively, which are used to extract the final key. Then the classical-classical-quantum state of Alice, Bob and Eve is given by
\begin{equation}
\rho_{\mathbf{X}^{n}\bar{\mathbf{X}}^{n}E}= \sum_{x^{n}, \bar{x}^{n}}\mathbb{P}(x^{n}, \bar{x} ^{n})
                                                                      \ket{x^{n}, \bar{x}^{n}}_{AB}\bra{x^{n}, \bar{x}^{n}} \otimes \rho_{E|{x^{n}, \bar{x}^{n}}},   
\end{equation}
where $x^{n} \in \mathbf{X}^{n}$ and $\bar{x}^{n} \in \bar{\mathbf{X}}^{n}$ respectively, and $\mathbb{P}(x^{n}, \bar{x}^{n})$ is the probability of joint bit string ($x^{n}, \bar{x}^{n}$). Similarly, we define a bit string $a^{n} \in A$ that is given by $a^{n}=x^{n} \oplus \bar{x}^{n}$, where the plus (modular 2) is bitwise (it is a XOR operation on bits). Then we classify the state $\rho_{\mathbf{X}^{n}\bar{\mathbf{X}}^{n}E}$ according to the bit string $a^{n}$, and define the conditional state
\begin{equation}
\begin{aligned}
\rho_{\mathbf{X}^{n}\bar{\mathbf{X}}^{n}E|a^{n}}:= \sum_{x^{n} \oplus \bar{x}^{n}=a^{n}} \frac{\mathbb{P}(x^{n}, \bar{x}^{n})}{\sum_{x^{n} \oplus \bar{x}^{n}=a^{n}}\mathbb{P}(x^{n}, \bar{x}^{n})}   \ket{x^{n}, \bar{x}^{n}}_{AB}\bra{x^{n}, \bar{x}^{n}} \otimes \rho_{E|x^{n}, \bar{x}^{n}},   
\end{aligned}
\end{equation}
and its corresponding probability by $\mathbb{P}(a^{n})$ which is a marginal probability distribution $\mathbb{P}(x^{n}, \bar{x}^{n})$. Therefore, $\rho_{\mathbf{X}^{n} \bar{\mathbf{X}}^{n}E}$ can be rewritten by $\rho_{\mathbf{X}^{n} \bar{\mathbf{X}}^{n}E}= \sum_{a^{n}} \mathbb{P}(a^{n}) \rho_{\mathbf{X}^{n} \bar{\mathbf{X}}^{n}E|a^{n}}$. For a conditional state $\rho_{\mathbf{X}^{n} \bar{\mathbf{X}}^{n}E|a^{n}}$, we consider two POVMs. One is the $\mathbb{X}^{n}$,  the other one is $\mathbb{M}^{\otimes n} (a^{n})$ given by
\begin{equation}
\begin{aligned}
\mathbb{M}^{\otimes n} (a^{n}):=\mathbb{M}_{a_{1}} \otimes \cdots \otimes\mathbb{M}_{a_{i}} \otimes \cdots \otimes\mathbb{M}_{ a_{n}}. 
\end{aligned}
\end{equation}
where $\mathbb{M}_{0}=\mathbb{Y}$ and $\mathbb{M}_{1}=\mathbb{Z}$, $a_{i} \in \{0, 1\}$ is exactly the $i_{\text{th}}$ bit of the bit string $a^{n}$ (for example, if $a^{n}=00 \cdots 10 \cdots 11$, then $\mathbb{M}^{\otimes n} (00 \cdots 10 \cdots 11)=\mathbb{Y} \otimes\mathbb{Y} \otimes \cdots \otimes\mathbb{Z} \otimes \mathbb{Y}  \otimes \cdots \otimes \mathbb{Z}\otimes\mathbb{Z}$). Then, we use the uncertainty relation for smooth entropies \cite{tomamichel2011uncertainty}. For any tripartite quantum state $\rho_{ABE} \in \mathcal{H}_{ABE}$, if $\mathbf{M}^{n}$ and $\bar{\mathbf{M}}^{n}$ are the outcome bit strings after applying $\mathbb{M}^{\otimes n} (a^{n})$ to Alice and Bob's quantum system respectively, and the post-measurement state is $\rho_{\mathbf{M}^{n} \bar{\mathbf{M}}^{n}}$,then 
\begin{equation}
   H_{\text{min}}^{\bar{\varepsilon}}(\mathbf{X}^{n}|E)_{\rho_{\mathbf{X}^{n}E|a^{n}}}
+ H_{\text{max}}^{\bar{\varepsilon}}(\mathbf{M}^{n} | \bar{\mathbf{M}}^{n} )_{\rho_{\mathbf{M}^{n} \bar{\mathbf{M}}^{n}}} \ge n,
\end{equation}
where $\rho_{\mathbf{X}^{n}E|a^{n}}=\text{Tr}_{\text{B}} (\rho_{\mathbf{X}^{n}\bar{\mathbf{X}}^{n}E|a^{n}})$.

\emph{Proof.} According to \cite{tomamichel2011uncertainty}, we have 
\begin{equation}
   H_{\text{min}}^{\bar{\varepsilon}}(\mathbf{X}^{n}|E)_{\rho_{\mathbf{X}^{n}E|a^{n}}}
+ H_{\text{max}}^{\bar{\varepsilon}}(\mathbf{M}^{n} | B )_{\rho_{\mathbf{M}^{n} B}} \ge \log_{2} \frac{1}{c}=n,
\end{equation}
where we have use the fact that, for each $a^{n}$, the overlap of $\mathbb{X}^{\otimes n}$ and $\mathbb{M}^{\otimes n} (a^{n})$ is $1/2^n$, and the post-measurement state $\rho_{\mathbf{M}^{n} B}$ is obtained after Alice applies $\mathbb{M}^{\otimes n} (a^{n})$ to her particle.
Considering the data-processing inequality for smooth max-entropy \cite{tomamichel2010duality}, we have 
\begin{equation}
H_{\text{max}}^{\bar{\varepsilon}}(\mathbf{M}^{n} | B )_{\rho_{\mathbf{M}^{n} B}}
\le H_{\text{max}}^{\bar{\varepsilon}}(\mathbf{M}^{n} | \bar{\mathbf{M}}^{n} )_{\rho_{\mathbf{M}^{n} \bar{\mathbf{M}}^{n}}},
\end{equation}
which we complete the proof.

As we know, there are $2^n$ bit strings in the set $A$, therefore, there are $2^n$ smooth min-entropies. To connect the smooth min-entropy for $\rho_{\mathbf{X}^{n}E}=\text{Tr}_{\text{B}}(\rho_{\mathbf{X}^{n}\bar{\mathbf{X}}^{n}E})$ and $2^n$ smooth min-entropies for each $\rho_{\mathbf{X}^{n}E|a^{n}}$,  we introduce the following lemma.


\hfill

\noindent {\bf Lemma 1.} For any normalized density matrix $\rho=\sum_{i=1}p_{i}\rho_{i}$ with the constraint $\sum_{i=1}p_{i}=1$, if there exists an unnormalized density matrix $\tilde\rho=\sum_{i=1}\tilde p_{i}\tilde\rho_{i}$ that satisfy $P(p_{i}, \tilde p_{i})\le \varepsilon$ and $\text{Max}_{i} P(\rho_{i}, \tilde\rho_{i})\le \bar{\varepsilon}$ where $P(\cdot , \cdot)$ denotes purified distance \cite{tomamichel2010duality}, then 
\begin{equation}
P(\rho, \tilde\rho)\le \sqrt{1-(1-\varepsilon^{2})(1-\bar{\varepsilon}^{2})}.
\end{equation}

\emph{Proof.} Because of the constraint $\sum_{i=1}p_{i}=1$, we find that $\rho_{i}$ is normalized. According to the definition of the purified distance (see Section 2 in \cite{tomamichel2010duality}), we have 
\begin{equation}
P(\rho_{i}, \tilde\rho_{i})=\sqrt{1-\bar {F}^{2}(\rho_{i}, \tilde\rho_{i})}=\sqrt{1-F^{2}(\rho_{i}, \tilde\rho_{i})} \le \bar{\varepsilon},
\end{equation}
where $\bar{F}(\cdot , \cdot)$ denotes purified fidelity. Owing to the strong concavity of the fidelity, we find that 
\begin{equation}
\label{constext}
\begin{aligned}
F(\rho, \tilde\rho)&\ge \sum_{i=1}\sqrt{p_{i} \tilde p_{i}} F(\rho_{i}, \tilde\rho_{i}) 
                             \ge \sum_{i=1}\sqrt{p_{i} \tilde p_{i}} \sqrt{1-\bar{\varepsilon}^{2}} 
                            =\sqrt{1-\bar{\varepsilon}^{2}}\sum_{i=1}\sqrt{p_{i} \tilde p_{i}} 
                            =\sqrt{1-\varepsilon^{2}}\sqrt{1-\bar{\varepsilon}^{2}}. \\
\end{aligned}
\end{equation}
Thus, we have 
\begin{equation}
\begin{aligned}
P(\rho, \tilde\rho)= \sqrt{1-\bar F^{2}(\rho, \tilde\rho)} 
                           \le \sqrt{1-(1-\varepsilon^{2})(1-\bar{\varepsilon}^{2})}, 
\end{aligned}
\end{equation}
which completes the proof.


\hfill

\noindent {\bf Lemma 2.}  For a normalized density matrix $\rho_{AB}=\sum_{i=1}p_{i}\rho_{i}$ with the constraint $\sum_{i=1}p_{i}=1$ and a unnormalized density matrix $\tilde\rho_{AB}=\sum_{i=1}\tilde p_{i}\tilde\rho_{i}$ that satisfy $P(p_{i}, \tilde p_{i})\le \varepsilon$ and $H_{\text{min}}(A|B)_{\tilde\rho_{i}}=H_{\text{min}}^{\bar{\varepsilon}}(A|B)_{\rho_{i}}$ for each index $i$, we have
\begin{equation}
2^{-H_{\text{min}}^{\sqrt{1-(1-\varepsilon^{2})(1-\bar{\varepsilon}^{2})}}(A|B)_{\rho}}\le \sum_{i} \tilde p_{i} 2^{-H_{\text{min}}^{\bar{\varepsilon}}(A|B)_{\rho_{i}}}.
\end{equation}

\emph{Proof.} $H_{\text{min}}(A|B)_{\tilde\rho_{i}}=H_{\text{min}}^{\bar{\varepsilon}}(A|B)_{\rho_{i}}$ suggests that $\text{Max}_{i} P(\rho_{i}, \tilde\rho_{i})\le \bar{\varepsilon}$. Combining Lemma 1, we have $P(\rho_{AB}, \tilde\rho_{AB})\le \sqrt{1-(1-\varepsilon^{2})(1-\bar{\varepsilon}^{2})}$. Thus, we obtain 
\begin{equation}
H_{\text{min}}^{\sqrt{1-(1-\varepsilon^{2})(1-\bar{\varepsilon}^{2})}}(A|B)_{\rho} \ge H_{\text{min}}(A|B)_{\tilde \rho}.
\end{equation}
Owing to the sub-additivity of min-entropy, we have 
\begin{equation}
2^{-H_{\text{min}}(A|B)_{\tilde \rho}} \le \sum_{i} \tilde p_{i} 2^{-H_{\text{min}}(A|B)_{\tilde \rho_{i}}}.
\end{equation}
Consequently, 
\begin{equation}
2^{-H_{\text{min}}^{\sqrt{1-(1-\varepsilon^{2})(1-\bar{\varepsilon}^{2})}}(A|B)_{\rho}} 
\le 2^{-H_{\text{min}}(A|B)_{\tilde \rho}} 
\le \sum_{i} \tilde p_{i} 2^{-H_{\text{min}}(A|B)_{\tilde \rho_{i}}}
=  \sum_{i} \tilde p_{i} 2^{-H_{\text{min}}^{\bar{\varepsilon}}(A|B)_{\rho_{i}}} .
\end{equation}
which completes the proof.

Owing to Lemma 2, we gain some intuition that the lower bound of smooth min-entropy for a "big" state $\rho$ can be obtained by the summation of some "small" states $\rho_{i}$ if $\rho=\sum_{i=1}p_{i}\rho_{i}$. In this section, the "big" state is $\rho_{\mathbf{X}^{n}E}$, the "small" states are $\rho_{\mathbf{X}^{n}E|a^{n}}$, and they satisfy $\rho_{\mathbf{X}^{n}E}=\sum_{a^{n}}\mathbb{P}(a^{n}) \rho_{\mathbf{X}^{n}E|a^{n}}$. 


Then, for the length by $n$ bit string $a^{n}$, we denote the frequency distribution by $\gamma_{t}$ of "$t$", which is defined by the relative number of occurrences of each "$t$", that is 
\begin{equation}
\gamma_{t}:=\frac{1}{n}|\{i: a_{i}=t\}|,
\end{equation}
for any $t \in \{0, 1\}$. Actually, if we do not consider statistical fluctuation, $\gamma_{1}=Q_x$, for $Q_x$ is error bit rate calculated from an $m$-tuple of elements sampled at random from the $(n+m)$-tuple of elements in the $\mathbb{X}$-measurement and $\gamma_{1}$ is the frequency distribution of the leftover $n$-tuple of elements. Therefore, if we exclude a small probability denoted by $\varepsilon^{2}$ event and only consider its mutually exclusive event that the error rate under $\mathbb{X}$-basis measurement is bounded by $Q_x +\mu$, then we can find a probability distribution 
\begin{equation}
\label{eq6}
\mathbb{Q}(a^{n}) :=\left\{
\begin{aligned}
\frac{\mathbb{P}(a^{n})}{1-\varepsilon^{2}} & , &  \gamma_{1} \le Q_x +\mu, \\
0 & , & \text{else}.
\end{aligned}
\right.
\end{equation}
Thus, we can find that $F(\mathbb{P}, \mathbb{Q})=\sum_{a^{n}} \sqrt{{\mathbb{P}(a^{n})}\mathbb{Q}(a^{n})}= \sqrt{1-\varepsilon^{2}}$ and then the purified distance \cite{tomamichel2010duality} between the distributions is given by $P(\mathbb{P}, \mathbb{Q})=\sqrt{1-F^{2}(\mathbb{P}, \mathbb{Q})}=\varepsilon$.

In the following, we focus on bounding $H_{\text{max}}^{\bar{\varepsilon}}(\mathbf{M}^{n} | \bar{\mathbf{M}}^{n} )_{\rho_{\mathbf{M}^{n} \bar{\mathbf{M}}^{n}}}$ by the observed values. Firstly, we note that the correlation of $\mathbf{M}^{n}$ and $\bar{\mathbf{M}}^{n}$ is discussed when Alice and Bob respectively output bit strings $\mathbf{M}^{n}$ and $\bar{\mathbf{M}}^{n}$ that satisfy $a^{n}=x^{n} \oplus \bar{x}^{n} $. Then we can conceive a hypothetical experiment that, if we already know the outputs are $x^{n}$ on Alice's side and $\bar{x}^{n}$ on Bob's side under $\mathbb{X}^{\otimes n}$-basis measurement, which satisfy $a^{n}=x^{n} \oplus \bar{x}^{n}$, we do the POVM $\mathbb{M}^{\otimes n} (a^{n})$ and record the output values. Finally, the output values in the hypothetical experiment help us analyze the correlation of $\mathbf{M}^{n}$ and $\bar{\mathbf{M}}^{n}$ and thus bound $H_{\text{max}}^{\bar{\varepsilon}}(\mathbf{M}^{n} | \bar{\mathbf{M}}^{n} )_{\rho_{\mathbf{M}^{n} \bar{\mathbf{M}}^{n}}}$. In fact, we can use the actual observed values to reconstruct the values that we need in the hypothetical experiment up to a failure probability. 

For this purpose, we recall that it is sufficient to consider that the quantum states shared by Alice and Bob before any measurements have the simple form \cite{renner2005information} 
\begin{equation}
\rho^{n}_{AB}=\sum^{n}_{n_{00}, n_{01}, n_{10}, n_{11}} \mu_{n_{00}, n_{01}, n_{10}, n_{11}} \rho^{n}_{n_{00}, n_{01}, n_{10}, n_{11}}.
\end{equation}
Similarly to equation (1) in \cite{renner2005information}, the sum is taken over all $n_{00}, n_{01}, n_{10}, n_{11} \in \mathbb{N}_{0}$ satisfying $n_{00}+n_{01}+n_{10}+n_{11}=n$ and $ \mu_{n_{00}, n_{01}, n_{10}, n_{11}}$ are some non-negative coefficients. Moreover, $\rho^{n}_{n_{00}, n_{01}, n_{10}, n_{11}}$ is the state of n qubit pairs defined by
\begin{equation}
\rho^{n}_{n_{00}, n_{01}, n_{10}, n_{11}}:=\pi_{n}((\ket{\Phi_{00}}\bra{\Phi_{00}})^{\otimes n_{00}}\otimes (\ket{\Phi_{01}}\bra{\Phi_{01}})^{\otimes n_{01}} \otimes (\ket{\Phi_{10}}\bra{\Phi_{10}})^{\otimes n_{10}}\otimes (\ket{\Phi_{11}}\bra{\Phi_{11}})^{\otimes n_{11}}),
\end{equation}
where the operator $\pi_{n}$ denotes the completely positive map which symmetries the state with respect to permutations of the n qubit pairs and $\ket{\Phi_{00}}:=1/\sqrt{2}(\ket{00}+\ket{11})$, $\ket{\Phi_{01}}:=1/\sqrt{2}(\ket{01}+\ket{10})$, $\ket{\Phi_{10}}:=1/\sqrt{2}(\ket{00}-\ket{11})$, $\ket{\Phi_{11}}:=1/\sqrt{2}(\ket{01}-\ket{10})$ are the Bell states. Then we define the frequency distributions that 
\begin{equation}
\begin{aligned}
&\lambda_{00}:=\frac{1}{n}\sum^{n}_{n_{00}, n_{01}, n_{10}, n_{11}} \mu_{n_{00}, n_{01}, n_{10}, n_{11}} n_{00},   \quad   \text{for}  \ket{\Phi_{00}}\bra{\Phi_{00}} \\
&\lambda_{01}:=\frac{1}{n}\sum^{n}_{n_{00}, n_{01}, n_{10}, n_{11}} \mu_{n_{00}, n_{01}, n_{10}, n_{11}} n_{01},   \quad   \text{for}  \ket{\Phi_{01}}\bra{\Phi_{01}} \\
&\lambda_{10}:=\frac{1}{n}\sum^{n}_{n_{00}, n_{01}, n_{10}, n_{11}} \mu_{n_{00}, n_{01}, n_{10}, n_{11}} n_{10},   \quad   \text{for}  \ket{\Phi_{10}}\bra{\Phi_{10}} \\
&\lambda_{11}:=\frac{1}{n}\sum^{n}_{n_{00}, n_{01}, n_{10}, n_{11}} \mu_{n_{00}, n_{01}, n_{10}, n_{11}} n_{11},   \quad   \text{for}  \ket{\Phi_{11}}\bra{\Phi_{11}}, \\
\end{aligned}
\end{equation}
where these frequency distributions satisfy $\lambda_{00}+\lambda_{01}+\lambda_{10}+\lambda_{11}=1$.
Additionally, we find $\ket{\Phi_{00}}$ outcomes no error regardless of applying $\mathbb{X}$-measurement, $\mathbb{Y}$-measurement or $\mathbb{Z}$-measurement. $\ket{\Phi_{01}}$ outcomes an error when applying $\mathbb{Y}$-measurement and $\mathbb{Z}$-measurement. $\ket{\Phi_{10}}$ outcomes an error when applying $\mathbb{X}$-measurement and $\mathbb{Z}$-measurement. $\ket{\Phi_{11}}$ outcomes an error when applying $\mathbb{X}$-measurement and $\mathbb{Y}$-measurement. Therefore, we have $Q_x=\lambda_{10}+\lambda_{11}$, $Q_y=\lambda_{01}+\lambda_{11}$ and $Q_z=\lambda_{01}+\lambda_{10}$, or, equivalently, $\lambda_{00}=1-\frac{1}{2}(Q_x+Q_y+Q_z)$, $\lambda_{01}=\frac{1}{2}(-Q_x+Q_y+Q_z)$, $\lambda_{10}=\frac{1}{2}(Q_x-Q_y+Q_z)$ and $\lambda_{11}=\frac{1}{2}(Q_x+Q_y-Q_z)$. Consequently, as we required in the hypothetical experiment picture, we can define the "conditional" value denoted by 
\begin{equation}
\xi_{0|0}:=\frac{1-\frac{1}{2}(Q_x+Q_y+Q_z)}{1-Q_x},
\end{equation}
which accounts for, if we already know that the frequency distribution of "0" of the bit string $a^{n}$ is $1-Q_x$, the frequency distribution of no error after both Alice and Bob apply the POVM $\mathbb{M}^{\otimes n} (a^{n})$. Then, to analyze the correlation of $\mathbf{M}^{n}$ and $\bar{\mathbf{M}}^{n}$, we consider the probability distribution $\mathbb{R}(b^{n})$ of a bit string $b^{n}:=m^{n} \oplus \bar{m}^{n} $, where $m^{n} \in \mathbf{M}^{n}$, $\bar{m}^{n} \in \bar{\mathbf{M}}^{n}$ and $b^{n} \in B$. Similarly, for a bit string $b^{n}$ the $i_{\text{th}}$ bit of which is denoted by $l_i$, we denote a "conditional" frequency distribution by $\beta_{j}|\gamma_{t}$, that is,
\begin{equation}
\beta_{j}|\gamma_{t}:=\frac{1}{n \times \gamma_{t}}|\{i: l_{i}=j \land j_{i}=t\}|,
\end{equation}
for any $j \in \{0, 1\}$. Similarly, if we do not consider statistical fluctuation, $\beta_{0}|\gamma_{0}=\xi_{0|0}$. Then, if we exclude a small probability denoted by $1-(1-\varepsilon^{2})^{2}$ event and only consider its mutually exclusive event that the error rate under $\mathbb{Y}$-basis measurement is bounded by $Q_y+\mu$ and $\mathbb{Z}$-basis measurement is bounded by $Q_z+\mu$, then we can find a probability distribution
\begin{equation}
\label{eq6}
\mathbb{S}(b^{n}):=\left\{
\begin{aligned}
\frac{\mathbb{R}(b^{n})}{(1-\varepsilon^{2})^{2}} & , &\beta_{0}|\gamma_{0} \ge \frac{1-\frac{1}{2}(Q_x+Q_y+Q_z+3\mu)}{1-Q_x-\mu},\\
0 & , & \text{else}.
\end{aligned}
\right.
\end{equation}
Similarly, we find that $F(\mathbb{R}, \mathbb{S})=\sum_{b^{n}} \sqrt{\mathbb{S}(b^{n})  \mathbb{R}(b^{n})}=\sqrt{(1-\varepsilon^{2})^{2}}$. Then the purified distance between the distributions is given by $P(\mathbb{R}, \mathbb{S})=\sqrt{1-F^{2}(\mathbb{R}, \mathbb{S})}=\sqrt{1-(1-\varepsilon^{2})^{2}}$. Hence,  with the assumption $Q_x=Q_y=Q_z=Q$, under the distribution $\mathbb{S}$, the total number of errors on $n(1-Q-\mu)$ bits (from $\mathbb{Y}$-measurement) is at most $W_y:= \lfloor n(1-Q-\mu)(1-\frac{1-\frac{3}{2}(Q+\mu)}{1-Q-\mu}) \rfloor$, similarly, the total number of errors on $n(Q+\mu)$ bits (from $\mathbb{Z}$-measurement) is at most $W_z:= {\lfloor n(Q+\mu)\frac{\frac{1}{2}(Q+\mu)}{Q+\mu} \rfloor} \le \frac{1}{2}n(Q+\mu)$. Owing to the technique in \cite{tomamichel2012tight} (see Lemma 3 in Supplementary Information), we have 
\begin{equation}
\begin{aligned}
    & H_{\text{max}}^{\bar{\varepsilon}}(\mathbf{M}^{n} | \bar{\mathbf{M}}^{n} )_{\mathbb{R}} 
 \le  H_{\text{max}}(\mathbf{M}^{n} | \bar{\mathbf{M}}^{n} )_{\mathbb{S}}  \\
 \le & \log_{2}(\sum_{w_y}^{W_y} \dbinom{n(1-Q-\mu)}{w_y} \times \sum_{w_z}^{W_z}\dbinom{n(Q+\mu)}{w_y}) 
 \le  n(1-Q-\mu)H_{2}(\frac{1-\frac{3}{2}(Q+\mu)}{1-Q-\mu}) +n(Q+\mu),  \\
\end{aligned}
\end{equation}
where $\bar{\varepsilon}:=\sqrt{1-(1-\varepsilon^{2})^{2}}$ and $H_{2}(\cdot)$ denotes the binary entropy function. Hence, we can obtain the lower bound of $H_{\text{min}}^{\tilde{\varepsilon}}(\mathbf{X}^{n}|E)_{\rho_{\mathbf{X}^{n}E}}$, given by
\begin{equation}
H_{\text{min}}^{\tilde{\varepsilon}}(\mathbf{X}^{n}|E)_{\rho_{\mathbf{X}^{n}E}}
 \ge n[1-H_{2}(\frac{1-\frac{3}{2}(Q+\mu)}{1-Q-\mu})][1-Q-\mu],
 \end{equation}
where $\tilde{\varepsilon}:=\sqrt{1-(1-\varepsilon^{2})(1-\bar{\varepsilon}^{2})}=\sqrt{1-(1-\varepsilon^{2})^{3}}$.

\emph{Proof.} For $\rho_{\mathbf{X}^{n}E}=\sum_{a^{n}}\mathbb{P}(a^{n}) \rho_{\mathbf{X}^{n}E|a^{n}}$, we have $H_{\text{min}}^{\tilde{\varepsilon}}(\mathbf{X}^{n}|E)_{\rho_{\mathbf{X}^{n}E}} \ge  -\log_{2} \sum_{a^{n}} \mathbb{Q}(a^{n}) 2^{-H_{\text{min}}^{\bar{\varepsilon}}(\mathbf{X}^{n}|E)_{\rho_{\mathbf{X}^{n}E|a^{n}}}}$ according to Lemma 2. For each $\rho_{\mathbf{X}^{n}E|a^{n}}$, its smooth min-entropy satisfies 
\begin{equation}
\begin{aligned}
H_{\text{min}}^{\bar{\varepsilon}}(\mathbf{X}^{n}|E)_{\rho_{\mathbf{X}^{n}E|a^{n}}} 
\ge n-
H_{\text{max}}^{\bar{\varepsilon}}(\mathbf{M}^{n} | \bar{\mathbf{M}}^{n} )_{\rho_{\mathbf{M}^{n} \bar{\mathbf{M}}^{n}}}.
\end{aligned}
\end{equation}
Thus, we have
\begin{equation}
\begin{aligned}
     H_{\text{min}}^{\tilde{\varepsilon}}(\mathbf{X}^{n}|E)_{\rho_{\mathbf{X}^{n}E}}  
     \ge & -\log_{2} \sum_{a^{n}} \mathbb{Q}(a^{n}) 2^{-H_{\text{min}}^{\bar{\varepsilon}}(\mathbf{X}^{n}|E)_{\rho_{\mathbf{X}^{n}E|a^{n}}}}  \\
\ge & -\log_{2} \sum_{a^{n}} \mathbb{Q}(a^{n}) 2^{-n+n[(1-Q-\mu) H_{2}(\frac{1-\frac{3}{2}(Q+\mu)}{1-Q-\mu})+(Q+\mu) ]}  \\
=    & -\log_{2} 2^{-n+n[(1-Q-\mu) H_{2}(\frac{1-\frac{3}{2}(Q+\mu)}{1-Q-\mu})+(Q+\mu)]}  \\
=    & n[1-H_{2}(\frac{1-\frac{3}{2}(Q+\mu)}{1-Q-\mu})][1-Q-\mu], \\
\end{aligned}
\end{equation}
which completes the proof. Owing to the Quantum Leftover Hashing Lemma, we finally obtain Eq (A.16).


\section*{appendix b: full security proof}
Following the idea introduced in Appendix A, we present the full proof of our main result for generalized case. 


\hfill

If $\mathbf{X}_{1, \vec{j}}^{n} $ and $\bar{\mathbf{X}}_{1, \vec{j}}^{n}$ are the outcome dit strings after applying $\mathbb{X}_{1, \vec{j}}^{\otimes n} (y_{0,1}^{n})$ to Alice and Bob's quantum system respectively, and the post-measurement state is $\rho_{\mathbf{X}_{1, \vec{j}}^{n} \bar{\mathbf{X}}_{1, \vec{j}}^{n}}$,then 
\begin{equation}
   H_{\text{min}}^{\bar{\varepsilon}}(\mathbf{X}_{0,1}^{n}|E)_{\rho_{\mathbf{X}_{0,1}^{n}E|y_{0,1}^{n}}}
+ H_{\text{max}}^{\bar{\varepsilon}}(\mathbf{X}_{1, \vec{j}}^{n} | \bar{\mathbf{X}}_{1, \vec{j}}^{n} )_{\rho_{\mathbf{X}_{1, \vec{j}}^{n} \bar{\mathbf{X}}_{1, \vec{j}}^{n}}}
\ge n\log_{2}d,
\end{equation}

\emph{Proof.} In the main text, we have obtained that
\begin{equation}
  H_{\text{min}}^{\bar{\varepsilon}}(\mathbf{X}_{0,1}^{n}|E)_{\rho_{\mathbf{X}_{0,1}^{n}E|y_{0,1}^{n}}}
+H_{\text{max}}^{\bar{\varepsilon}}(\mathbf{X}_{1, \vec{j}}^{n} |B)_{\rho_{\mathbf{X}_{1, \vec{j}}^{n}B}} 
\ge n\log_{2}d. 
\end{equation}
Considering the data-processing inequality for smooth max-entropy \cite{tomamichel2010duality}, we have 
\begin{equation}
H_{\text{max}}^{\bar{\varepsilon}}(\mathbf{X}_{1, \vec{j}}^{n} |B)_{\rho_{\mathbf{X}_{1, \vec{j}}^{n}B}}
\le H_{\text{max}}^{\bar{\varepsilon}}(\mathbf{X}_{1, \vec{j}}^{n} | \bar{\mathbf{X}}_{1, \vec{j}}^{n} )_{\rho_{\mathbf{X}_{1, \vec{j}}^{n} \bar{\mathbf{X}}_{1, \vec{j}}^{n}}},
\end{equation}
which completes the proof.


\hfill

\noindent {\bf Definition 3.} In main text, we have defined the $i_{th}$ dit of string $y_{0,1}^{n}$ by $j_{i} \in \{0, \cdots ,d-1\}$. Thus, for the length by $n$ dit string $y_{0,1}^{n}$, we denote the frequency distribution by $\gamma_{t}$ of "$t$", which is defined by the relative number of occurrences of each "$t$", that is 
\begin{equation}
\gamma_{t}:=\frac{1}{n}|\{i: j_{i}=t\}|,
\end{equation}
for any $t \in \{0, \cdots ,d-1\}$. Actually, if we do not consider statistical fluctuation, $\gamma_{t}=q^{(t)}_{0,1}$, for $q^{(t)}_{0,1}$ is calculated from an $m$-tuple of elements sampled at random from the $(n+m)$-tuple of elements in the $\mathbb{X}_{0,1}$-measurement and $\gamma_{t}$ is the frequency distribution of the leftover $n$-tuple of elements. Therefore, if we exclude a small probability denoted by $\varepsilon^{2}$ event and only consider its mutually exclusive event that the error rate under $\mathbb{X}_{0,1}$-basis measurement is bounded by $1-q^{(0)}_{0,1}+\mu$, then we can find a probability distribution 
\begin{equation}
\label{eq6}
\mathbb{Q}(y_{0,1}^{n}) :=\left\{
\begin{aligned}
\frac{\mathbb{P}(y_{0,1}^{n})}{1-\varepsilon^{2}} & , & \gamma_{0} \ge q^{(0)}_{0,1}-\mu, \\
0 & , & \text{else}.
\end{aligned}
\right.
\end{equation}
Thus, we can find that $F(\mathbb{P}, \mathbb{Q})=\sum_{y_{0,1}^{n}} \sqrt{{\mathbb{P}(y_{0,1}^{n})}\mathbb{Q}(y_{0,1}^{n})}= \sqrt{1-\varepsilon^{2}}$. Then the purified distance \cite{tomamichel2010duality} between the distributions is given by $P(\mathbb{P}, \mathbb{Q})=\sqrt{1-F^{2}(\mathbb{P}, \mathbb{Q})}=\varepsilon$.


In the following, we focus on bounding $H_{\text{max}}^{\bar{\varepsilon}}(\mathbf{X}_{1, \vec{j}}^{n} | \bar{\mathbf{X}}_{1, \vec{j}}^{n})_{\rho_{\mathbf{X}_{1, \vec{j}}^{n} \bar{\mathbf{X}}_{1, \vec{j}}^{n}}}$ by observed values. First, we note that the correlation of $\mathbf{X}_{1, \vec{j}}^{n}$ and $\bar{\mathbf{X}}_{1, \vec{j}}^{n}$ is discussed when Alice and Bob respectively output dit strings $\mathbf{X}_{0,1}^{n}$ and $\bar{\mathbf{X}}_{0,1}^{n}$ that satisfy $y_{0,1}^{n}=\bar{x}_{0,1}^{n}-x_{0,1}^{n} \pmod{d} $. Then we can conceive a hypothetical experiment that, if we already know the outputs are $x_{0,1}^{n}$ on Alice's side and $\bar{x}_{0,1}^{n}$ on Bob's side under $\mathbb{X}^{\otimes n}_{0,1}$-basis measurement, which satisfy $y_{0,1}^{n}=\bar{x}_{0,1}^{n}-x_{0,1}^{n} \pmod{d}$, we do the POVM $\mathbb{X}_{1, \vec{j}}^{\otimes n} (y_{0,1}^{n})$ and record the output values. Finally, the output values in the hypothetical experiment help us analyze the correlation of $\mathbf{X}_{1, \vec{j}}^{n}$ and $\bar{\mathbf{X}}_{1, \vec{j}}^{n}$ and thus bound $H_{\text{max}}^{\bar{\varepsilon}}(\mathbf{X}_{1, \vec{j}}^{n} | \bar{\mathbf{X}}_{1, \vec{j}}^{n})_{\rho_{\mathbf{X}_{1, \vec{j}}^{n} \bar{\mathbf{X}}_{1, \vec{j}}^{n}}}$. In fact, we can use the actual observed values to reconstruct the values that we need in the hypothetical experiment up to a failure probability. 

For this purpose, we recall that it is sufficient to consider that the quantum states shared by Alice and Bob before any measurements have the simple form \cite{renner2005information} 
\begin{equation}
\begin{aligned}
&\rho^{n}_{AB}=\sum^{n}_{n_{00}, \cdots, n_{jk}, \cdots, n_{d-1,d-1}} \mu_{n_{00}, \cdots, n_{jk}, \cdots, n_{d-1,d-1}} \rho^{n}_{n_{00}, \cdots, n_{jk}, \cdots, n_{d-1,d-1}}.\\
\end{aligned}
\end{equation}
In this formula, the sum is taken over all $n_{00}, \cdots, n_{jk}, \cdots, n_{d-1,d-1} \in \mathbb{N}_{0}$ satisfying $\sum_{j, k=0}^{d-1}n_{jk}=n$ and $\mu_{n_{00}, \cdots, n_{jk}, \cdots, n_{d-1,d-1}}$ are some non-negative coefficients. Moreover, there exists a unitary operation $\pi_n$ on $\mathcal{H}^{n}_{AB}$ which permutes the $n$ subsystems, so that the $n$ qudit pairs $\rho^{n}_{n_{00}, \cdots, n_{jk}, \cdots, n_{d-1,d-1}}$ can be given by 
\begin{equation}
\begin{aligned}
&\rho^{n}_{n_{00}, \cdots, n_{jk}, \cdots, n_{d-1,d-1}}:=\pi_{n}(\otimes _{j, k=0}^{d-1}(\ket{\Phi_{jk}}\bra{\Phi_{jk}})^{\otimes n_{jk}}).\\
\end{aligned}
\end{equation}
In this expression, the generalized high-dimensional Bell basis states $\ket{\Phi_{jk}}=\sum_{s=0}^{d-1}\omega^{sk}\ket{s,s+j}$ ($j, k\in\{0, 1, \cdots, d-1\}$ and $\omega$ is the $d$th root of unity) \cite{sheridan2010security} belong to the composed Hilbert space of Alice and Bob denoted by $\mathcal{H}_{AB}$.
We note that the pair of qudits $\ket{\Phi_{t,kt-j \bmod d}}$ outcomes "$t$" type "error" under measurements of $\mathbb{X}_{0,1}$-basis and "$j$" type "error" under measurements of $\mathbb{X}_{1,k}$-basis, and we define its corresponding expected value 
\begin{equation}
\begin{aligned}
&\lambda_{t,kt-j \bmod d}:=\frac{1}{n}\sum^{n}_{n_{00},\cdots, n_{d-1,d-1}} \mu_{n_{00}, \cdots, n_{d-1,d-1}} n_{t,kt-j \bmod d}.\\
\end{aligned}
\end{equation}
Then, we connect the actual observed values with $\lambda_{t,kt-j \bmod d}$ that
\begin{equation}
\begin{aligned}
&q_{0,1}^{(t)}=\sum_{kt-j \bmod d}\lambda_{t,kt-j \bmod d}  \qquad  q_{1,k}^{(t)}=\sum_{t}\lambda_{t,kt-j \bmod d}, \\
\end{aligned}
\end{equation}
or, equivalently, 
\begin{equation}
\lambda_{t,kt-j \bmod d}=\frac{1}{d}(\sum_{s} q_{1,s}^{((s-k)t+j \bmod d)} + q_{0,1}^{(t)} -1).
\end{equation}
Consequently, we can define the "conditional" values as we required in the hypothetical experiment picture. These values are given by
\begin{equation}
\xi_{j|t}:=\frac{\lambda_{t,t^{2}-j \bmod d}}{q_{0,1}^{(t)}} ,
\end{equation}
which account for the expected probability that $\rho^{n}_{n_{00}, \cdots, n_{jk}, \cdots, n_{d-1,d-1}}$ outcomes "$j$" type "error" in $\mathbb{X}_{1,t}$-basis under the condition that this pair of qudits outcomes "$t$" type "error" in $\mathbb{X}_{0,1}$-basis.


\hfill

\noindent {\bf Definition 4.} Here, we consider a probability distribution with respect to the pair of dit strings $(\mathbf{X}_{1, \vec{j}}^{n}, \bar{\mathbf{X}}_{1, \vec{j}}^{n})$ that is denoted by $\mathbb{R}(x_{1, \vec{j}}^{n}, \bar{x}_{1, \vec{j}}^{n})$. For each pair of dit strings $(x_{1, \vec{j}}^{n}, \bar{x}_{1, \vec{j}}^{n})$, we find a dit string $z_{1, \vec{j}}^{n} \in \mathbf{Z}_{1, \vec{j}}^{n}$ by doing subtraction bitwise, that is $z_{1, \vec{j}}^{n}=\bar{x}_{1, \vec{j}}^{n}-x_{1, \vec{j}}^{n}   \pmod{d}$. Then, we define a marginal probability denoted by $\mathbb{R}(z_{1, \vec{j}}^{n})$ of $\mathbb{R}(x_{1, \vec{j}}^{n}, \bar{x}_{1, \vec{j}}^{n})$. Similarly to Definition 3, for a dit string $z_{1, \vec{j}}^{n}(y_{0,1}^n)$ the $i_{\text{th}}$ dit of which is denoted by $l_i$ and the corresponding dit string $y_{0,1}^n$ of which has a frequency distribution $\gamma_t$, we denote a conditional frequency distribution by $\beta_{j}|\gamma_{t}$, that is,
\begin{equation}
\beta_{j}|\gamma_{t}:=\frac{1}{n \times \gamma_{t}}|\{i: l_{i}=j \land j_{i}=t\}|,
\end{equation}
for any $j \in \{0, \cdots ,d-1\}$. Similarly, if we do not consider statistical fluctuation, $\beta_{j}|\gamma_{t}=\xi_{j|t}$. Then, if we exclude a small probability denoted by $1-(1-\varepsilon^{2})^{d}$ event and only consider its mutually exclusive event that the error rate under all $\mathbb{X}_{1,t}$-basis measurement is bounded by $1-q_{1,t}^{(0)}+\mu$, then we can find a probability distribution
\begin{equation}
\label{eq6}
\mathbb{S}(z_{1, \vec{j}}^{n}(y_{0,1}^n)):=\left\{
\begin{aligned}
\frac{\mathbb{R}(z_{1, \vec{j}}^{n}(y_{0,1}^n))}{(1-\varepsilon^{2})^{d}} & , &\beta_{0}|\gamma_{t} \ge \xi_{0|t}(q^{(0)}_{j,k}-\mu), \  \text{for all} \ t\\
0 & , & \text{else},
\end{aligned}
\right.
\end{equation}
where $\xi_{0|t}$ is a function of $q^{(0)}_{j,k}-\mu$ as equation (A.36) and (A.37) show, and the subscript $(j,k) \in \{(0,1), (1,0), \cdots, (1,d-1)\}$. Similarly, we find that $F(\mathbb{R}, \mathbb{S})=\sum_{z_{1, \vec{j}}^{n}(y_{0,1}^n)} \sqrt{\mathbb{S}(z_{1, \vec{j}}^{n}(y_{0,1}^n))  \mathbb{R}(z_{1, \vec{j}}^{n}(y_{0,1}^n))}=\sqrt{(1-\varepsilon^{2})^{d}}$. Then the purified distance between the distributions is given by $P(\mathbb{R}, \mathbb{S})=\sqrt{1-F^{2}(\mathbb{R}, \mathbb{S})}=\sqrt{1-(1-\varepsilon^{2})^{d}}$. 


\hfill

\noindent {\bf Lemma 5.} We define a probability vector
\begin{equation}
\underline{\xi}_{0}=\{\xi_{0|0}, \xi_{1|0}, \cdots, \xi_{d-1|0}\}, \quad
\text{where} \ \xi_{0|0}=\frac{d(1-Q-\mu)+q_{0,1}^{(0)}-1} {d \times q_{0,1}^{(0)}} \ \text{and} \ \xi_{1|0}= \cdots= \xi_{d-1|0}=\frac{1-\xi_{0|0}}{d-1},
\end{equation}
and let $\bar{\varepsilon}=\sqrt{1-(1-\varepsilon^{2})^{d}}$.
Then
\begin{equation}
\begin{aligned}
   H_{\text{max}}^{\bar{\varepsilon}}(\mathbf{X}_{1, \vec{j}}^{n} | \bar{\mathbf{X}}_{1, \vec{j}}^{n} )_{\rho_{\mathbf{X}_{1, \vec{j}}^{n} \bar{\mathbf{X}}_{1, \vec{j}}^{n}}}
 \le n[\gamma_{0}H(\underline{\xi}_{0})+(1-\gamma_{0}) \log_{2} d].
\end{aligned}
\end{equation}

\emph{Proof.} Owing to the definition of smooth max-entropy and the technique introduced in \cite{tomamichel2012tight}, we have
\begin{equation}
\begin{aligned}
      &   H_{\text{max}}^{\bar{\varepsilon}}(\mathbf{X}_{1, \vec{j}}^{n} | \bar{\mathbf{X}}_{1, \vec{j}}^{n} )_{\mathbb{R}} 
 \le     H_{\text{max}}   (\mathbf{X}_{1, \vec{j}}^{n} | \bar{\mathbf{X}}_{1, \vec{j}}^{n} )_{\mathbb{S}}  \\
 \le &\log_{2}\sum_{\omega_{1|0}, \cdots, \omega_{d-1|0}=0}^{\omega_{1|0}+\cdots+\omega_{d-1|0}\le \lfloor n \times \gamma_{0} \times (1-\beta_{0}|\gamma_{0}) \rfloor}   \frac{(n \times \gamma_{0})!}{\omega_{0|0}! \ \omega_{1|0}! \cdots \omega_{d-1|0}!} 
 \times \prod_{t=1}^{d-1}\sum_{\omega_{0|t}, \cdots, \omega_{d-1|t}=0}   \frac{(n \times \gamma_{t})!}{\omega_{0|t}! \cdots \omega_{d-1|t}!}  \\
 =  &\log_{2} \sum_{\omega_{1|0}, \cdots, \omega_{d-1|0}=0}^{\omega_{1|0}+\cdots+\omega_{d-1|0}\le \lfloor n \times \gamma_{0} \times (1-\beta_{0}|\gamma_{0}) \rfloor}   \frac{(n \times \gamma_{0})!}{\omega_{0|0}! \ \omega_{1|0}! \cdots \omega_{d-1|0}!} 
 +  \sum_{t=1}^{d-1} \log_{2} \sum_{\omega_{0|t}, \cdots, \omega_{d-1|t}=0}   \frac{(n \times \gamma_{t})!}{\omega_{0|t}! \cdots \omega_{d-1|t}!} \\
 \le & n[\gamma_{0}H(\underline{\xi}_{0})+(1-\gamma_{0}) \log_{2} d],  \\
\end{aligned}
\end{equation}
where we have used the assumption that the quantum channel is a generalization of the qubit depolarizing channel (that is, $q_{1,k}^{(0)}=1-Q$ for each $k$), and the fact that $\beta_{0}|\gamma_{0} \ge \xi_{0|0}$. The last inequality is shown in Lemma 6.


\hfill

\noindent {\bf Lemma 6.} Let $\sum_{j=0}^{d-1}\Omega_{j}=N$ with $\Omega_{0}\ge N/2$, and definite the corresponding probability vector $\underline{\Omega}:= \{\Omega_{0}/N, \Omega_{1}/N, \cdots, \Omega_{d-1}/N\} $. Then
\begin{equation}
\sum_{\omega_{1}=0}^{\Omega_{1}} \cdots \sum_{\omega_{d-1}=0}^{\Omega_{d-1}}
\frac{N!}{\omega_{0}! \ \omega_{1}! \cdots \omega_{d-1}!} \le 2^{N*H(\underline{\Omega})}, \qquad
\text{where} \  \sum_{j=0}^{d-1}\omega_{j}=N.
\end{equation}

\emph{Proof.} Combining the facts that 
\begin{equation}
\begin{aligned}
1&=(\sum_{j=0}^{d-1}\frac{\Omega_{j}}{N})^{N}=\sum_{\omega_{0}, \cdots, \omega_{d-1}} \frac{N!}{\omega_{0}! \  \cdots \omega_{d-1}!} (\frac{\Omega_{0}}{N})^{\omega_{0}} \cdots (\frac{\Omega_{d-1}}{N})^{\omega_{d-1}} \\
&\ge \sum_{\omega_{1}=0}^{\Omega_{1}} \cdots \sum_{\omega_{d-1}=0}^{\Omega_{d-1}}
\frac{N!}{\omega_{0}! \cdots \omega_{d-1}!} (\frac{\Omega_{0}}{N})^{\omega_{0}} \cdots (\frac{\Omega_{d-1}}{N})^{\omega_{d-1}}, \\
\end{aligned}
\end{equation}
and 
\begin{equation}
\begin{aligned}
(\frac{\Omega_{0}}{N})^{\omega_{0}} \cdots (\frac{\Omega_{d-1}}{N})^{\omega_{d-1}}
\ge (\frac{\Omega_{0}}{N})^{\Omega_{0}} \cdots (\frac{\Omega_{d-1}}{N})^{\Omega_{d-1}}, 
\end{aligned}
\end{equation}
we obtain
\begin{equation}
\begin{aligned}
1& \ge (\frac{\Omega_{0}}{N})^{\Omega_{0}} \cdots (\frac{\Omega_{d-1}}{N})^{\Omega_{d-1}}  \sum_{\omega_{1}=0}^{\Omega_{1}} \cdots \sum_{\omega_{d-1}=0}^{\Omega_{d-1}} \frac{N!}{\omega_{0}! \cdots \omega_{d-1}!} \\
  &=2^{\sum_{j=0}^{d-1}\Omega_{j}\log_{2}\frac{\Omega_{j}}{N}} \sum_{\omega_{1}=0}^{\Omega_{1}} \cdots \sum_{\omega_{d-1}=0}^{\Omega_{d-1}} \frac{N!}{\omega_{0}! \cdots \omega_{d-1}!} 
  =2^{-N*H(\underline{\Omega})} \sum_{\omega_{1}=0}^{\Omega_{1}} \cdots \sum_{\omega_{d-1}=0}^{\Omega_{d-1}} \frac{N!}{\omega_{0}! \cdots \omega_{d-1}!}. \\
\end{aligned}
\end{equation}
It remains to prove equation (A.61). We figure that $\Omega_{0}\ge N/2 \ge \text{Max}\{ \Omega_{1}, \cdots, \Omega_{d-1}\}$, thus 
\begin{equation}
\begin{aligned}
\Omega_{0}^{\omega_{0}-\Omega_{0}} \ge (\text{Max}\{ \Omega_{1}, \cdots, \Omega_{d-1}\})^{\sum_{j=1} ^{d-1}\Omega_{j}-\omega_{j}} = \prod_{j=1} ^{d-1} (\text{Max}\{ \Omega_{1}, \cdots, \Omega_{d-1}\})^{\Omega_{j}-\omega_{j}} \ge \prod_{j=1} ^{d-1}  \Omega_{j}^{\Omega_{j}-\omega_{j}}.
\end{aligned}
\end{equation}
Equivalently, $\prod_{j=0} ^{d-1}\Omega_{j}^{\omega_{j}}\ge\prod_{j=0} ^{d-1}\Omega_{j}^{\Omega_{j}}$ which completes the proof of equation (A.61), and consequently we complete the full proof of Lemma 6. Moreover, $\Omega_{0}/N \ge \xi_{0|0}$, $H(\underline{\Omega})$ reaches maximum when $\Omega_{1}/N=\cdots=\Omega_{d-1}/N=\frac{1-\xi_{0|t}}{d-1}$, that is, $H(\underline{\Omega})\le H(\underline{\xi}_{0})$. Additionally, it is not hard to note that
\begin{equation}
\begin{aligned}
d^{n_{t}}=(1+\cdots+1)^{n_{t}}=\sum_{\omega_{0|t}, \cdots, \omega_{d-1|t}=0}   \frac{n_{t}!}{\omega_{0|t}! \cdots \omega_{d-1|t}!}.
\end{aligned}
\end{equation}

Now that all ingredients are ready, we prove our result as follows

\hfill

\noindent {\bf Theorem 7.}  Let $\tilde{\varepsilon}:=\sqrt{1-(1-\varepsilon^{2})(1-\bar{\varepsilon}^{2})}=\sqrt{1-(1-\varepsilon^{2})^{d+1}}$, then
\begin{equation}
\begin{aligned}
H_{\text{min}}^{\tilde{\varepsilon}}(\mathbf{X}_{0,1}^{n}|E)_{\rho_{\mathbf{X}_{0,1}^{n}E}}
 \ge n(\log_{2}d-H({\underline{\xi}}))(1-Q-\mu).
\end{aligned}
\end{equation}

\emph{Proof.} For $\rho_{\mathbf{X}_{0,1}^{n}E}=\sum_{y_{0,1}^{n}}\mathbb{P}(y_{0,1}^{n}) \rho_{\mathbf{X}_{0,1}^{n}E|y_{0,1}^{n}}$, owing to Lemma 2, we have 
\begin{equation}
\begin{aligned}
H_{\text{min}}^{\sqrt{1-(1-\varepsilon^{2})(1-\bar{\varepsilon}^{2})}}(\mathbf{X}_{0,1}^{n}|E)_{\rho_{\mathbf{X}_{0,1}^{n}E}}
\ge  -\log_{2} \sum_{y_{0,1}^{n}} \mathbb{Q}(y_{0,1}^{n}) 2^{-H_{\text{min}}^{\bar{\varepsilon}}(\mathbf{X}_{0,1}^{n}|E)_{\rho_{\mathbf{X}_{0,1}^{n}E|y_{0,1}^{n}}}} . 
\end{aligned}
\end{equation}
For each $\rho_{\mathbf{X}_{0,1}^{n}E|y_{0,1}^{n}}$, its smooth min-entropy satisfies 
\begin{equation}
\begin{aligned}
H_{\text{min}}^{\bar{\varepsilon}}(\mathbf{X}_{0,1}^{n}|E)_{\rho_{\mathbf{X}_{0,1}^{n}E|y_{0,1}^{n}}} \ge n\log_{2} d-
H_{\text{max}}^{\bar{\varepsilon}}(\mathbf{X}_{1, \vec{j}}^{n} | \bar{\mathbf{X}}_{1, \vec{j}}^{n} )_{\rho_{\mathbf{X}_{1, \vec{j}}^{n} \bar{\mathbf{X}}_{1, \vec{j}}^{n}}}.
\end{aligned}
\end{equation}
Combining Lemma 5 and the probability distribution $\mathbb{Q}$ where $\mathbb{Q}(y_{0,1}^{n})=0$ when $\gamma_{0} \ge q_{0,1}^{(0)}-\mu = 1-Q-\mu$, we obtain
\begin{equation}
\begin{aligned}
      H_{\text{max}}^{\bar{\varepsilon}}(\mathbf{X}_{1, \vec{j}}^{n} | \bar{\mathbf{X}}_{1, \vec{j}}^{n} )_{\rho_{\mathbf{X}_{1, \vec{j}}^{n} \bar{\mathbf{X}}_{1, \vec{j}}^{n}}} 
 \le & n[\gamma_{0}H(\underline{\xi}_{0})+(1-\gamma_{0}) \log_{2} d] \\
 \le & n[(1-Q-\mu) H(\underline{\xi}_{0})+(Q+\mu) \log_{2} d] \\
 \le & n[(1-Q-\mu) H(\underline{\xi})+(Q+\mu) \log_{2} d] \\
\end{aligned}
\end{equation}
where we use the facts $q_{0,1}^{(0)}=1-Q$ and $H(\underline{\xi}_{0})\le H(\underline{\xi})\le \log_{2}d$. We finally obtain that 
\begin{equation}
\begin{aligned}
     H_{\text{min}}^{\tilde{\varepsilon}}(\mathbf{X}_{0,1}^{n}|E)_{\rho_{\mathbf{X}_{0,1}^{n}E}}  
     \ge & -\log_{2} \sum_{y_{0,1}^{n}} \mathbb{Q}(y_{0,1}^{n}) 2^{-H_{\text{min}}^{\bar{\varepsilon}}(\mathbf{X}_{0,1}^{n}|E)_{\rho_{\mathbf{X}_{0,1}^{n}E|y_{0,1}^{n}}}}  \\
\ge & -\log_{2} \sum_{y_{0,1}^{n}} \mathbb{Q}(y_{0,1}^{n}) 2^{-n\log_{2}d+n[(1-Q-\mu) H(\underline{\xi})+(Q+\mu) \log_{2} d]}  \\
=    & -\log_{2} 2^{-n\log_{2}d+n[(1-Q-\mu) H(\underline{\xi})+(Q+\mu) \log_{2} d]}  \\
=    & n(\log_{2}d-H({\underline{\xi}}))(1-Q-\mu), \\
\end{aligned}
\end{equation}
which completes the proof.

\hfill

\noindent {\bf Theorem 8.} The (d+1)-basis protocols $\Phi[n, m, l, \varepsilon_{cor}, leak_{EC}]$ using $d$-level quantum states is $\varepsilon_{sec}$-secret for some $\varepsilon_{sec}>0$ if $l$ satisfies 
\begin{equation}
\begin{aligned}
 l \le \mathop{\text{max}}\limits_{\tilde{\varepsilon},\varepsilon^{'}}   \lfloor n(\log_{2}d-H({\underline{\xi}}))(1-Q-\mu(\varepsilon))-2\log_{2} \frac{1}{2\varepsilon^{'}}-leak_{EC}-\log_{2}\frac{2}{\varepsilon_{cor}} \rfloor,    
\end{aligned}
\end{equation}
where 
\begin{equation}
\begin{aligned}
H({\underline{\xi}})=-\frac{1-\frac{d+1}{d}(Q+\mu)}{1-Q-\mu)} \log_{2} \frac{1-\frac{d+1}{d}(Q+\mu)}{1-Q-\mu}-(d-1) \frac{\frac{1}{d(d-1)}(Q+\mu)}{1-Q-\mu} \log_{2} \frac{\frac{1}{d(d-1)}(Q+\mu)}{1-Q-\mu},
\end{aligned}
\end{equation}
and we optimize over $\varepsilon>0$ and $\varepsilon^{'}>0$ with constraints
\begin{equation}
\begin{aligned}
\tilde{\varepsilon}+\varepsilon^{'}\le \varepsilon_{sec},\ \tilde{\varepsilon}=\sqrt{1-(1-\varepsilon^{2})^{d+1}} \ \text{and} \   \mu(\varepsilon):=\sqrt{\frac{n+m}{nm}\frac{m+1}{m}\ln \frac{1}{\varepsilon}}. 
\end{aligned}
\end{equation}

\emph{Proof.} Due to the Quantum Leftover Hashing lemma \cite{renner2008security,tomamichel2011leftover}, it is possible to extract a $\Delta$-secret key of length $l$ from $\mathbf{X}_{0,1}^{n}$, where
\begin{equation}
\begin{aligned}
\Delta=2\tilde{\varepsilon}+\frac{1}{2}\sqrt{2^{l-H_{\text{min}}^{\tilde{\varepsilon}}(\mathbf{X}_{0,1}^{n}|E^{'})}}.
\end{aligned}
\end{equation}
The term $E^{'}$ that represents all information Eve obtained can be decomposed as $E^{'}=CE$, where $C$ is classical information revealed by Alice and Bob during the error correction step. For the revealed information that $C$ is at most $leak_{EC}+\log_{2}\frac{2}{\varepsilon_{cor}}$ bits, we use a chain rule for smooth entropies and then obtain
\begin{equation}
\begin{aligned}
H_{\text{min}}^{\tilde{\varepsilon}}(\mathbf{X}_{0,1}^{n}|E^{'}) \ge H_{\text{min}}^{\tilde{\varepsilon}}(\mathbf{X}_{0,1}^{n}|E)-leak_{EC}-\log_{2}\frac{2}{\varepsilon_{cor}}.
\end{aligned}
\end{equation}
With the lower bound of smooth min-entropy of $\rho_{\mathbf{X}_{0,1}^{n}E}$, we consequently get 
\begin{equation}
\begin{aligned}
\Delta \le 2\tilde{\varepsilon}+\frac{1}{2}\sqrt{2^{l-H_{\text{min}}^{\tilde{\varepsilon}}(\mathbf{X}_{0,1}^{n}|E^{'})}} \le 2\tilde{\varepsilon}+\varepsilon^{'} \le \varepsilon_{sec}.
\end{aligned}
\end{equation}
Thus, these protocols are $\varepsilon_{sec}$-secret.

\twocolumngrid

\nocite{*}

\bibliography{apssamp}

\end{document}